# Self-organizing Traffic Control: First Results

Carlos Gershenson
Centrum Leo Apostel, Vrije Universiteit Brussel
Krijgskundestraat 33, Brussels, 1160, Belgium
cgershen@vub.ac.be   http://homepages.vub.ac.be/~cgershen

**Abstract**. We developed a virtual laboratory for traffic control where agents use different strategies in order to self-organize on the road. We present our first results where we compare the performance and behaviour promoted by environmental constrains and five different simple strategies: three inspired in flocking behaviour, one selfish, and one inspired in the minority game. Experiments are presented for comparing the strategies. Different issues are discussed, such as the important role of environmental constrains and the emergence of traffic lanes.

## 1 Introduction

Self-organizing systems (Heylighen, 2003; Gershenson and Heylighen, 2003) can provide a useful approach for studying, designing, and controlling complex systems. In this paper we explore different self-organizing strategies used by agents to solve a traffic problem: there are n agents, sharing a resource, and they need to distribute their use of the resource in such a way that is "optimal" for most of them. In order to have a more specific environment, we focussed on automobile traffic control, but the ideas could be used for air traffic control, Internet traffic control, telecommunications traffic control, and similar tasks.

There have been a wide variety of studies on different aspects of automobile traffic control, and the role of self-organization on it (*e.g.* Nataksuji and Kaka, 1991; Chiu and Chand, 1993; Hoar, Penner, and Jacob, 2002). We will focus on a recent part of this research, namely when autonomous driving is involved (*e.g.* Gershenson, 1998; Kolodko and Vlacic, 2003).

For carrying out our investigations, we developed a virtual laboratory (Gershenson, González, and Negrete, 2000), which we named SOTraCon. It is available online (source code included) from http://homepages.vub.ac.be/~cgershen/sos.

In the next section, we state our position towards self-organization, which we consider a useful perspective for studying traffic control. In Section 3 we present the virtual laboratory in which we implemented different strategies, and in which we carried out the experiments exposed in Section 4. We discuss our results and observations in Section 5, for concluding and commenting future research directions in Section 6.

## 2 Self-organization: what do we mean?

The concept of self-organization has been used in a wide variety of contexts and with different meanings. For us, self organization is a way of describing systems, rather than a type of systems. This is because *any* dynamical system can be described as self-organizing or as self-disorganizing (Ashby, 1962; Gershenson and Heylighen, 2003). This is based on the tautology "any system tends to its more probable state". If it fits our purposes to call that most probable state organized, then the system can be said to be self-organizing. If we decide that the most probable state is disorganized, then we will speak of a self-disorganizing system. In spite of this, we believe that this approach is useful for understanding complex adaptive systems, since it allows us to observe and describe them at more than one level of abstraction (Gershenson, 2002a).

We call our traffic control self-organizing, because the global behaviour of the agents is a result of the individual behaviours and interactions. Again, any traffic can be said to be self-organizing, but the difference lies on how we observe and control the system: the control is not imposed, but implemented at the agent level in their strategies, for optimizing traffic at the global level. We observe and analyse the system at both levels.

We can also say that the traffic control is *emergent*, in the sense that the global behaviour cannot be deduced from the individual strategies of the agents. Rather, we have to observe this behaviour, for *then* trying to describe it in terms of the strategies.

## 3 SOTraCon: A Virtual Laboratory

Our virtual laboratory is being developed in Java using the Java 3D libraries. It consists of a virtual environment where agents drive through a straight cyclic road with bounded sides. The agents flow through the plane (z, x) of the space (x, y, z). The reader is invited to download the simulation in order to test it and to perceive better the traffic dynamics.

The user can navigate through the virtual space, observing agents using different strategies, and change and observe parameters. There are different updating schemes for studying the role of the updates in the outcome of simulations. The user can create and remove agents, which are generated at random positions (although this can later be specified). They have a time margin during which they are "immortal". After this, if agents collide between themselves, or against a side of the road, they crash, and stop being taken into account for the other agents (*i.e.* agents do not crash with crashed agents).

### 3.1 The Agents

The agents have a circular radius, generated randomly between 0.5 and 1.5 metres. The agents follow simple Newtonian dynamics to simulate their movement. They have a position (z, x), velocity (v), acceleration (a), and angle of orientation ($\theta$). According to these, their position is updated in the virtual environment each time step (equivalent to one second) with (1, 2):

$$v \mathrel{+}= a \qquad (1)$$

$$\begin{aligned} z &\mathrel{+}= v * \cos\theta \\ x &\mathrel{+}= v * \sin\theta \end{aligned} \qquad (2)$$

For the results presented in this paper, the updating is done asynchronously and random: each time step the agents are updated one by one, following a different random order.

The agents have a radius of perception (rp), which allows them to perceive other agents withing this range, in order to decide their actions. At each time step, all agents try to approach their cruise velocity increasing their acceleration according to (3), after initializing their acceleration with a=0:

$$a \mathrel{+}= \frac{v_{cruise} - v}{5 * v_{cruise}} \qquad (3)$$

They also try to go straight by adjusting their angle θ with (4):

$$\theta \leftarrow = 0.9 * \theta \qquad (4)$$

where the angle zero is straight, positive are to the left, and negative to the right.

Agents also try not to go over the side edges, by forcing the angle to be (5) if they are too close:

$$\theta = \frac{-x}{5 * width} \qquad (5)$$

If agents find one of their neighbours closer than a certain distance, proportional to their velocity, and also closer than a side margin, they make an "emergency break", forcing the acceleration to its minimum possible. This tries to prevent crashing with other agents, while at the same time being able to pass them with a certain side margin.

### 3.2 The Strategies

We tested the performance and self-organization of agents using five different strategies: three strategies inspired in flocking behaviour (Reynolds, 1987), a "selfish" strategy, and a strategy inspired in the minority game (Challet and Zhang, 1997).

The **first flocking strategy** calculates the centroid formed by the positions of the neighbours of the agent (closer than the radius of perception), which is calculated by averaging the position vectors of the neighbours. Then, the agent tries to go away from that centroid according to (6):

$$a \leftarrow = \frac{0.1 * rp}{z_{avg} - z}$$
$$\theta \leftarrow = \frac{x_{avg} - x}{width} \qquad (6)$$

where ($z_{avg}$, $x_{avg}$) is the vector of the centroid of the neighbours' positions. In other words, the agents try to keep as separated from the other agents as possible. The agents which are going very slow are not counted as neighbours. Otherwise, this creates strong oscillations in the velocities of all the agents.

The **second flocking strategy** also tries to go away from its neighbours using (6), but only of its closest ones, specifically the ones which are closer than rp/2. On the other hand, it tries to approach its neighbours closer than rp, including the close ones, in a similar way than for the avoidance (7):

$$a \leftarrow = \frac{0.1 * rp}{z_{avg} - z}$$
$$\theta \leftarrow = \frac{x_{avg} - x}{width} \qquad (7)$$

The avoidance part gives separation to the agents, while the approach gives them cohesion, similar to those observed in flocks, and modelled by Craig Reynolds (1987) with his famous *boids*.

The **third flocking strategy** is similar to the second, only that it tires to avoid the centroid of all its neighbours, using (6), while only trying to approach the close ones (rp/2), using (7).

The **selfish strategy** is very simple: it tries to pass the closest agent in front of it. To do so, it uses (8):

$$a+ = 0.5$$
$$\text{if } x_{closest} > 0 \text{ then } \theta- = 0.5 \quad (8)$$
$$\text{else } \theta+ = 0.5$$

which means that it accelerates, and turns to the right if the closest agent is in the left side of the road, and vice versa.

The **minority strategy** just counts how many of its neighbours are on which side of the road (left or right), and it turns to the left if there are more agents on the right side, and vice versa. If there are equal number of agents, the direction is not changed by the strategy.

## 4 Experiments

In order to observe and understand the performance delivered by different strategies, we devised different series of experiments on our virtual laboratory. We obtained statistics for comparing the performance of different strategies for different traffic densities, and for the performance of populations of mixed strategies. Statistical samples do not tell us the whole story, and thus we include descriptions of the agents' behaviours.

The detailed tables with the results of our experiments can be found at our website.

### 4.1 Performance with Different Traffic Densities

In order to observe the differences of the strategies under different traffic densities, we made five simulations for each case, during 5000 time steps (seconds), and measured the average velocity and standard deviation in each simulation. We used a length of the cyclic road of 100 metres, and a width of 9 metres. Agents were created with a rp of 30 metres and a random cruise velocity of 3±1 m/s. If all the agents have the same cruise velocity, then it is much easier to self-organize, all the strategies basically arrange the agents in space. The maximum allowed velocity was 5 m/s. We can appreciate the results in Figure 1.

When there is only one agent, the strategy does not make any difference: the agents just try to keep their cruise velocity. Since this is variable for different agents, this explains why there are differences in the graph. If to set cruise velocities exactly at a certain value, the agents approach this independently of their strategy.

When there are two agents, we can speak already about interactions. The flocking1 agents try to keep apart from each other, so when the faster agent tries to reach the slower one, the latter one tries to separate from the former, thus going faster, oscillating around the velocity of the faster agent, around a distance of rp. If the faster one gets too close, he will reduce his velocity, but because of inertia, the effect will be that both agents will be closer to the velocity of the faster, thus increasing the average velocity as compared with single agents. Flocking2 agents try to gather, but then to separate if they are too close, causing an oscillation around rp/2, which makes them to go slower on average. Flocking3 do not fall into this, because they try to separate on principle, similar to flocking1 agents. Since selfish agents try to pass the closest agent in front of them, when one passes the other, the other will try to pass the one, so they organize side by side, in an eternal race, going at maximum velocity. Minority agents just take their sides, and go at their cruise velocities. This is why there is a high standard deviation only in this strategy: in all the others the agents organize at similar velocities.

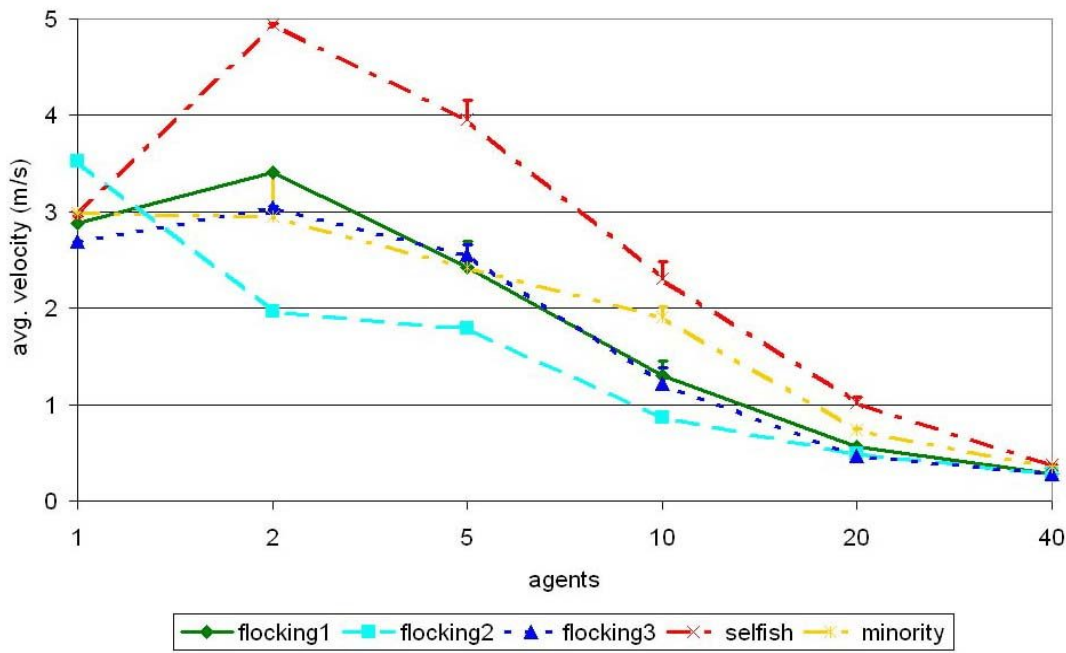

Figure 1. Average velocities for different agent densities.

When there are five agents in the simulation, flocking1 tend to oscillate a lot in their velocities, although different agents have different average velocities, *i.e.* travel different distances. This is similar for flocking3, but they do not oscillate so much, and going on average a bit faster. Flocking2 organize in a straight line, oscillating at a lower speed than the other strategies. Selfish agents organize very nicely, with a high variance meaning that agents go at different speeds, still organizing. It is a similar case with minority agents, which distribute themselves nicely, and faster agents can pass the slower ones.

With a population of ten agents, flocking1 agents tend to go to the sides of the road, but then they do not make passes, and oscillate a lot, which might cause some crashes. This is similar for flocking3, but with fewer oscillations, thus with fewer crashes. The standard deviations increase because each lane has an average velocity. Flocking2 agents try to stick together, but then the emergency breaks are activated, so that they keep a minimum distance, yet at a slow speed. Selfish agents self-organize to the edges of the road, but going fast since they are attempting to pass the agent which is ahead. Minority agents also form two lanes, each one going around the pace of the slowest agent. But for this concentration, which is not so bad, since the agents distribute themselves in the road almost optimally.

When there are 20 agents in the simulation, all the flocking agents keep their distances, but breaking every now and them. Flocking2 agents tend to leave less space between them, since they try to keep together. The selfish and minority strategies form two lanes, usually one faster than the other, and the agents drive a bit in zigzag. There are few passes.

With a density of 40 agents per 100 metres, it is difficult for the agents to move. Flocking agents keep their distances, but there is very little space, and this might cause some crashes. Selfish agents behave a bit chaotically, with several crashes (for the statistics, the agents were set as "immortal", otherwise soon there were only thirty agents left). Minority agents have the maximum separation of all, although still having few crashes. The crashes seem to be caused because with an asynchronous random updating, and agents being so close to each other, it might happen that one agent is updated twice before another. The updating intervals should be reduced to avoid this problem (*e.g.* 1 step=0.01s). It is worth noting that at this density, the strategy does not play a distinguishing role on the average velocities, only on the distributions of the agents, since they controlled much more by the environmental constrains (do not crash, do not go off the road) than by their strategies.

## 4.2 Dynamics Description

Even when all the strategies yield self-organization, in the sense that for random initial conditions, or during perturbations, the agents tend towards certain "preferred" configuration, this varies depending on the strategy. In this section we will describe these configurations for different strategies.

In Figure 2 we can appreciate images of groups of twenty agents implementing different strategies, and a combination of all five strategies. The flocking1 strategy, where agents try to separate from the centroid formed by their neighbours, makes the agents to keep a distance, both lateral and longitudinal. All the agents tend to go at the same velocity, since the faster ones push the slower ones, which try to get away from them, and the slower ones pull back the faster ones. Because of this, there is no passing. Flocking 2 agents tend to keep together, but if they are too close, they tend to separate. This causes an oscillation which propagates while agents switch between approaching and avoiding, so they tend to go slower because they have to break if others are breaking. Agents using a flocking3 strategy are more similar to flocking1, since they try to keep a distance from more agents than the ones they want to approach. Moreover, they are keeping a distance also from the agents which are close, so we could say that this strategy keeps a distance of the farther neighbours. This distributes the agents similarly than Flocking1. Selfish agents try to pass the agent in front of them, but when they do, this one will also try to pass them. This passing intention will cause agents to try to oscillate left and right, but when the density is high, they will not be able to do this if to avoid crashing, so they tend to keep their sides, forming emergent lanes. Minority agents go to both sides of the road, forming two lanes. Since usually one lane goes faster than the other, this can cause perturbations, in which some agents intend to change to the other side, if they perceive fewer agents on the other side (these perturbations decrease as the traffic density increases). If an agent is faster than the one in front of it, then taking advantage of the perturbations he will be able to pass. It is worth noting that minority agents are relatively stable independently if there are an even or odd number of them.

If we widen the road (*e.g.* to 15m), then selfish agents can form even three lanes, the middle one faster than the others. If the traffic is dense, the agents tend to oscillate more, making undulating lanes. Minority agents go to the sides, but perturbations still allow them to make some passes, which are frequent since there is more space between the two lanes formed by the agents. Flocking1 and Flocking3 agents tend to go towards the sides of the road, not taking advantage of the space left in the middle. Flocking2 agents form flocks, which go at the same velocity.

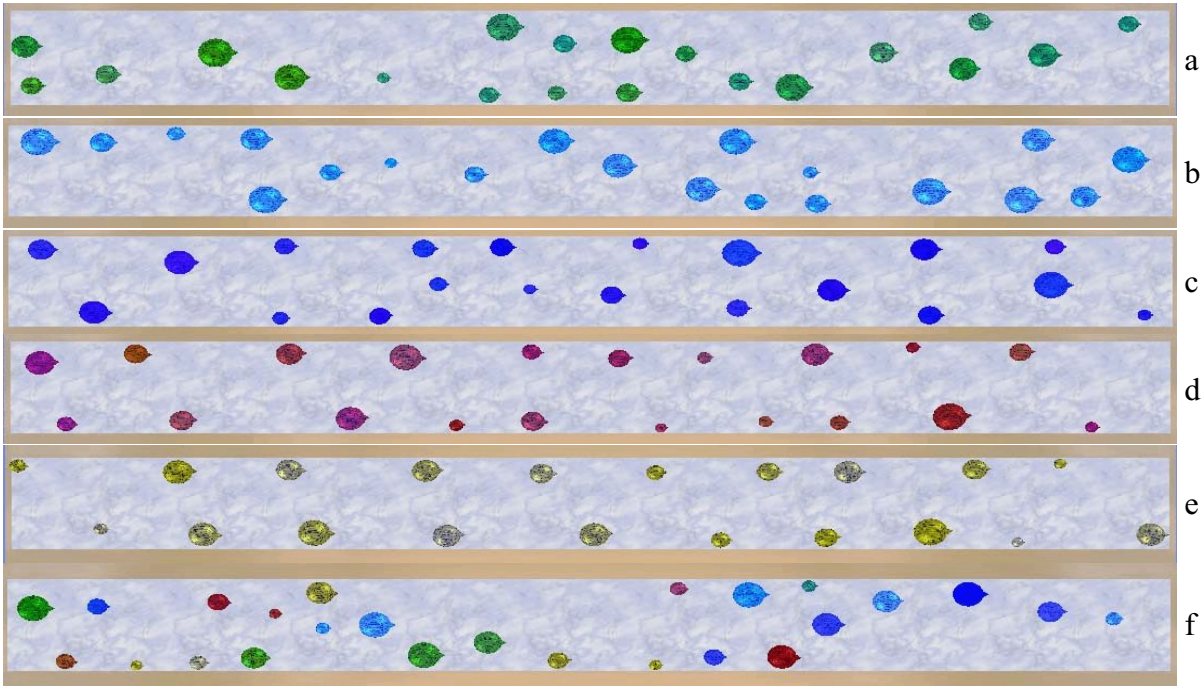

Figure 2. Patterns formed by twenty agents of different strategies: a) flocking1, b) flocking2, c) flocking3, d) selfish, e) minority, f) five agents of each strategy.

## 4.3 Mixing the strategies

If we set up agents of different strategies together, their behaviour will change, since we will have heterogeneous populations. In general, different strategies tend to block each other, so the agents cannot self-organize as well as with similar neighbours, and there is a higher probability of having crashes.

We made similar simulations than the ones described previously, with five agents of each type. The results can be appreciated in Figure 3.

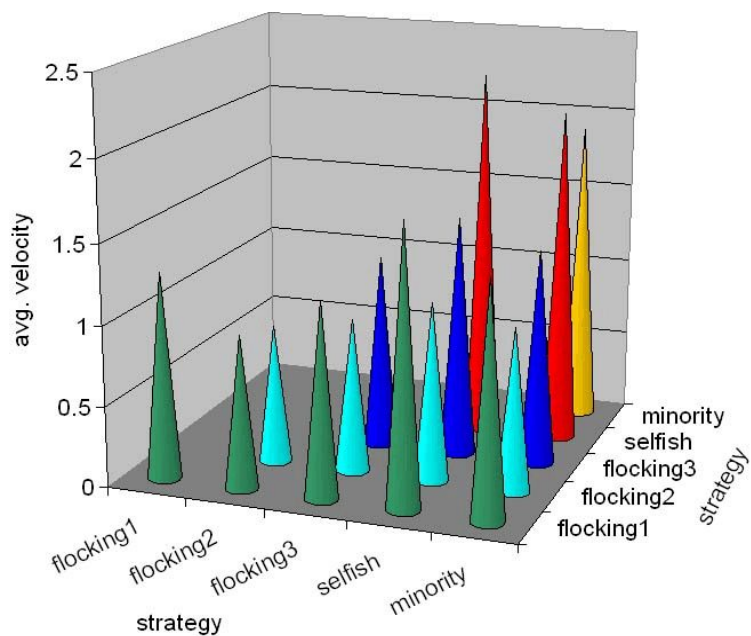

Figure 3. Average velocities for mixed populations, with five agents of each type.

The average velocity for mixed populations is between the averages of pure populations, in general tending to the slower one. The faster agents push the flocking strategies, and also these are pulled back by the slower ones. Minority agents are not pulled, but they do tend to push the flocking agents. Selfish agents try to go as fast as they can, so they push, but in some cases too much, causing crashes. In some cases the configuration does not allow them to pass anyway.

We also made tests where there would be nine agents of one strategy, and only one of another. In most cases the same principles hold: the average velocity is between the averages of both strategies, although tending more to the one of the large population. In some cases, however, the agent can even deteriorate a slower strategy, as is the case of a selfish agent in a flocking2 or minority population. In others, singe agents mix well, like a minority agent in a selfish population. In others, the agent can be benefited, as it is the case of flocking agents in a minority or selfish population, although the agent decreases the performance of the pure population.

## 5 Discussion

After observing the behaviour of the agents under different circumstances, one of the main things that we realized was the important role that the environmental constrains (do not crash, do not go off the road) play on the behaviour of the agents. When these constrains are too strong (very high traffic density), the strategy does not matter much, because the agent will try not to crash most of its time. This suggests that environmental constrains are an important component of a self-organizing process.

The flocking strategies tend to self-organize and adapt worse than the minority or selfish. The selfish one has the higher average velocities, but because of this, it also is the most prone to crash. In any case it is remarkable that they are able to self-organize. The flocking agents self-organize by all adopting a similar velocity, making few passes, whereas minority and selfish are able to adapt more and make passes more frequently. Flocking1 agents perform better under heavy traffic if their radius of perception is reduced, probably too many neighbours affect them. We should test a strategy similar to flocking1, but taking only into account the three closest neighbours.

Under different circumstances, and for different objectives, different strategies might be more suitable: If there is low traffic, selfish perform well, but they would waste more fuel than others. If the goal is to distribute better in space, minority and flocking1 can give different good alternatives. If the goal is that all agents should keep together, flocking2 is a good option with heavy traffic. This prompts us to explore mixed and also other more complex strategies.

It is worth noticing the emergence of driving lanes, generated by environmental constrains in all strategies except flocking2. If we would designate arbitrary lanes, this should improve the performance of the agents, because they would not get into each other's way. It is similar with human automobiles. If people would not follow lanes, traffic would be even more chaotic. It is also tempting to define not only arbitrary lanes, but dynamic virtual cells, which could be moving at the average velocities of their neighbours. Then agents would know if a cell is free if they would like to make a pass. This brings new considerations, such as conflict solving in case two or more agents would like to take a free cell at the same time.

Human drivers also self-organize, actually much better than these simple agents. The main idea behind autonomous driving, is that automobiles could communicate with each other, thus optimizing traffic flow. It seems that it would be too arrogant to waste the ability of a human driver, and the future of automobiles should include rather than only automated driving, a *collaborative* driving, where human and machine interact in order to optimize traffic flow. Machines can be used to communicate with other automobiles and provide information to the driver, who will be better at solving the uncertainties of the road. If someone claims that "intelligent" autonomous cars will solve all traffic problems, she does not know much about what she is speaking. Traffic jams are caused by the density of automobiles, and the best solution is to reduce the use of automobiles. Minor improvements are environmental constrains, such as barriers and transit codes, which optimize traffic flow. Good environmental constrains will make the silliest agent to self-organize in an optimal way, because there will be no other option left.

Our investigations would seem to be more relevant to traffic control in general than to particularly automobiles. Is it always the best alternative to try to use as much of the resources available? Not when there are in high, where it should be distributed to those who need it more. Self-organizing methods seem to be very useful tools for studying these situations. Simulations are required for understanding the emergent behaviour as agents interact among each other and with their environment. Understanding the effects at different levels of individual actions and global configurations can be done successfully from a self-organizational perspective.

# 6 Conclusions and Future Work

We presented the first results of a simple traffic simulation where agents follow simple strategies and according to these, different traffic patterns emerge. Further study is required for a) developing more strategies which will be able to cope better with a complex and variable environment; b) studying different environmental constrains, such as traffic lanes, which might improve traffic flow; and c) greater generalization of the lessons learned to be applied in general traffic control, and in self-organizing systems. Particularly, we are interested in studying the performance of different strategies in road crossings without traffic lights (Gershenson, 1998; Kolodko and Vlacic, 2003).

One relevant issue which should be also studied is the role of the updates in the simulation (synchronous, asynchronous, deterministic, random). It has become clear that depending on how we update the agents in a simulation, the behaviour of the system can change drastically (Gershenson, 2002b).

A related issue is the one of multicausality: if agents are affecting each other, and we do not know the exact order in which this takes place, we cannot describe the situation using classic causal models. We need to extend them, because in our simulations we saw clearly that A can cause certain effect in B, while at the same time B causes a certain effect in A. This is a general problem while modelling complex systems, when several elements interact with each other massively. Classical causality finds itself in "chicken and egg" problems, which we can come out using the appropriate philosophical tools. Simulations and virtual laboratories as the one we presented should be useful for developing these tools.

A final note for consideration is the role of the environment in control, especially of complex systems. It seems reasonable to design environments to constrain the behaviour of agents to solve a certain task, rather than trying to build intelligent agents able to cope with all the complexity an environment can give. But perfect environmental constrains can be as hard to build as perfect agents. Probably there should be a tradeoff: environmental constrains for making the task easier (reducing the environmental complexity), and agents able to adapt in this environment, but simple enough to allow their detailed study.

# Acknowledgements

I would like to thank Francis Heylighen for useful discussions on the subject. This work was supported in part by the Consejo Nacional de Ciencia y Tecnología (CONACYT) of México.